\documentclass[twocolumn,showpacs,preprintnumbers,amsmath,amssymb, prb, reprint]{revtex4-1}

\usepackage{stmaryrd}
\usepackage{txfonts}
\usepackage{amssymb}
\usepackage{mathrsfs}
\usepackage{graphicx}
\usepackage{dcolumn}
\usepackage{bm}
\usepackage{epsfig}
\usepackage{color}                    
\usepackage{hyperref}                 

\begin{document}
\preprint{\href{http://dx.doi.org/10.1063/1.4874677}{S.-Z. Lin, J. Appl. Phys. {\bf 115}, 173901 (2014).}}

\title{Mutual synchronization of two stacks of intrinsic Josephson junctions in cuprate superconductors}

\author{Shi-Zeng Lin}
\affiliation{Theoretical Division, Los Alamos National Laboratory, Los Alamos, New Mexico 87545, USA}

\begin{abstract}
Certain high-$T_c$ cuprate superconductors, 
which 
naturally realize a stack of Josephson junctions, thus can be used to generate electromagnetic waves in the terahertz region. A plate-like single crystal with $10^4$ junctions without cavity resonance was proposed to achieve strong radiation. For this purpose, it is required to synchronize the Josephson plasma oscillation in all junctions. In this work, we propose to use two stacks of junctions shunted in parallel to achieve synchronization. The two stacks are mutually synchronized in the whole IV curve, and there is a phase shift between the plasma oscillation in the two stacks. The phase shift is nonzero when the number of junctions in different stacks is the same, while it can be arbitrary when the number of junctions is different. This phase shift can be tuned continuously by applying a magnetic field when all the junctions are connected by superconducting wires. 
\end{abstract}
\date{\today}
\maketitle

\section{Introduction}
The Josephson effect predicted a half century ago can be used to convert an dc 
current 
input into high frequency electromagnetic (EM) oscillations. The angular frequency $\omega$ of the EM radiation can be tuned continuously by the 
dc voltage $V$ according to the ac Josephson relation $\omega=2 eV/\hbar$. The radiation from a single Josephson junction however is weak, of the order of pW  \cite{Yanson65,Dayem66,Zimmerma66} due to the large impedance mismatch \cite{Bulaevskii06PRL}. One natural way to enhance the radiation power is to fabricate arrays of Josephson junctions on a chip. \cite{Finnegan72,Jain84,Durala99,Barbara99,Song09} Once the arrays are fully synchronized, their coupling via the radiation field enhances and the radiation power becomes proportional  to the number of the junction squared, a phenomenon known as superradiation. In 1992 it was discovered that certain highly anisotropic cuprate superconductors, such as $\mathrm{Bi_2Sr_2CaCu_2O_{8+\delta}}$
(BSCCO) naturally realize a stack of intrinsic atomic-scale Josephson junctions (IJJs). \cite{Kleiner92} These junctions are homogeneous for a high quality crystal and they are coupled electromagnetically. 
Moreover because of the large superconducting energy gap, $\sim 60$ meV, the radiation frequency can be in the terahertz (THz) region, where the EM waves are promising for applications but are difficult to generate. 

Soon after the discovery of the IJJs effect, tremendous efforts have been made to excite strong THz radiation from cuprate superconductors, mainly BSCCO. \cite{Iguchi00b,Batov06,Bae07,Benseman11,Tachiki94,Koyama95,Tachiki05,Bulaevskii06,Bulaevskii07,szlin08,Koshelev08,szlin08b,Tachiki09,szlin09a,Wang09,Rakhmanov09,Wang10,Tsujimoto10,Tsujimoto12b} 
In 2007, strong and coherent radiation of THz EM waves was observed experimentally in a mesa structure of BSCCO with number of junctions about several hundred. 
\cite{Ozyuzer07} In such mesas with moderate number of junctions the Josephson plasma oscillations are synchronized by the cavity formed by the mesa itself, thus the radiation frequency is determined by the geometry of the mesa. For a rectangular mesa with dimension $L_x\times L_y\times L_z$, the wave vector of cavity modes is $\mathbf{k}=(n_x\pi/L_x,\ n_y\pi/L_y,\ n_z\pi/L_z)$. Typically $L_x\approx 50\ \mathrm{\mu m}$, $L_y\approx 300\ \mathrm{\mu m}$ and $L_z\approx 1\ \mathrm{\mu m}$. The in-phase modes with $n_z=0$ are responsible for the strong radiation. The radiation frequency is $f=c |k| /(2\pi\sqrt{\epsilon_d})$ with $\epsilon_d\approx 10$ being the dielectric constant of BSCCO. In most experiments, the mode $(n_x=1,\ n_y=0,\ n_z=0)$ was excited; while other high order modes with $n_z=0$ have also been observed recently. \cite{Kashiwagi2012} It is argued that it is the strong in-plane dissipation which is responsible for the selection of the $n_z=0$ mode. \cite{Koshelev10,szlin12a}  For a review on the recent progress, see Refs. \onlinecite{Hu10,Savelev10,Welp13}.

The observation of radiation from BSCCO crystal with the crystal working as a cavity (cavity design) has attracted considerable attention in the last several years for the possible high radiation power and extremely narrow linewidth \cite{LiMengyue12,Kashiwagi2012,szlin13b}. However the frequency tunability is limited in the cavity design. In 2007, an alternative design using a plate-like crystal was proposed. \cite{Bulaevskii07} In this plate design, the crystal along the $c$ axis is tall (typically $L_z\approx 40\ \mathrm{\mu m}$, thus involves $25000$ junctions). The lateral size is assumed to be about $L_x\times L_y\approx 4\times 300\ \mathrm{\mu m^2}$. Such $L_x$ is too short to support the cavity modes with frequency below the superconducting energy gap. It was shown that strong THz radiation can be achieved when the junctions are fully synchronized, with continuously tunable frequency since no cavity resonance is required in this design. The radiation power can be as high as $1$ mW with a relative intrinsic linewidth as narrow as $10^{-8}$. \cite{Bulaevskii11} It was shown that at such large number of junctions the in-phase oscillation of Josephson plasma can be stabilized by the radiation fields.  

At the initial stage after switching on the dc current, when IJJs are not synchronized, the radiation is weak thus it may be difficult to drive the system into the state with in-phase plasma oscillations, despite that the in-phase state with a strong radiation is stable. It was proposed then to synchronize a stack of IJJs with a shunt capacitor. \cite{szlin11,Bulaevskii11} By redistributing the current in the different branches of the circuit according to the phase dynamics of the IJJs, the capacitor can synchronize the plasma oscillations in the stack without the help of the radiation fields. The IJJs can also be synchronized by a shunting resistor or inductor, or by a more complicated shunt circuit. \cite{Hadley88} However, it is required that these shunting elements are capable of working in the THz band, which might be difficult to realize experimentally.

\begin{figure}[t]
\psfig{figure=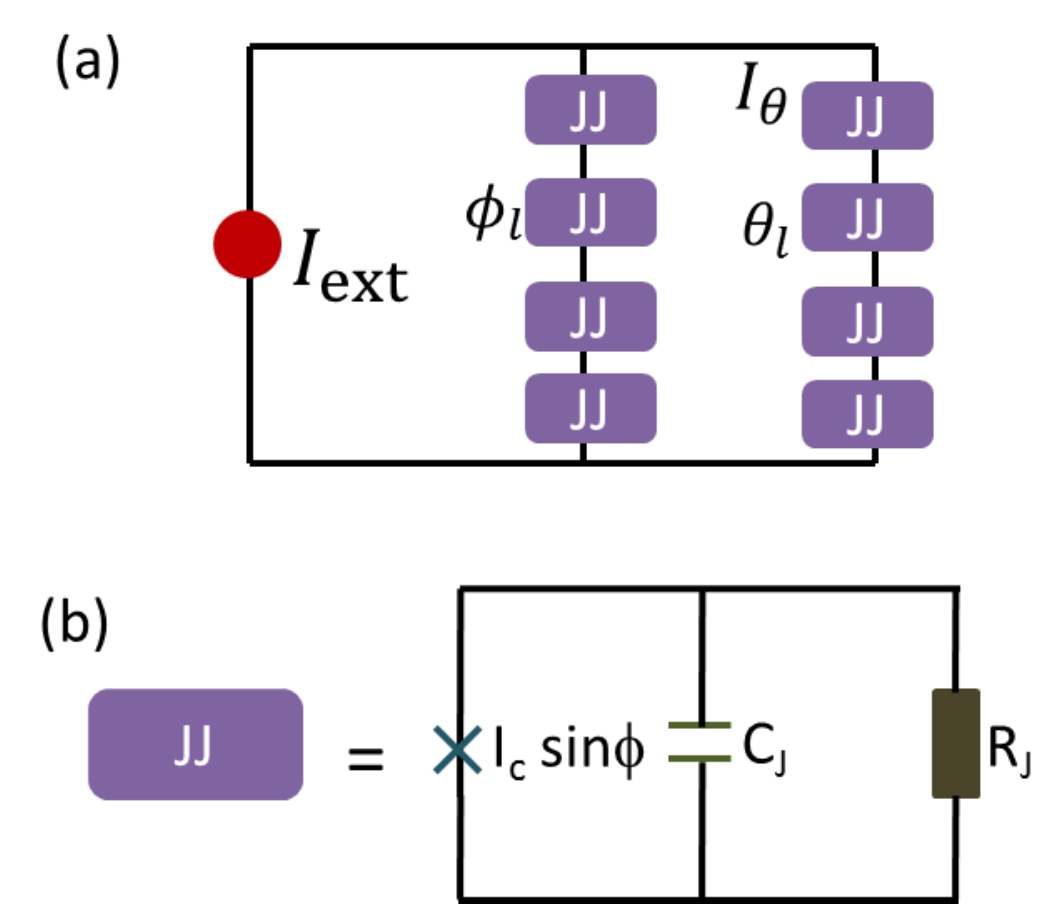,width=\columnwidth}
\caption{\label{f1}(color online) (a) Schematically view of two stacks of intrinsic Josephson junctions shunted in parallel and biased by a dc current. In order to tune the phase difference between the plasma oscillation in the two stacks by a magnetic field, the wire connecting all the junctions should also be superconducting. (b) The Josephson junction is modeled as a shunt circuit of a capacitor, a resistor, and a nonlinear Josephson current.}
\end{figure}

In this work, we propose to synchronize the Josephson plasma oscillations in junctions by shunting two stacks of IJJs in parallel. Since a Josephson junction can be regarded as a nonlinear inductor shunted by a resistor and a capacitor as depicted in Fig. \ref{f1}(b), it is expected the shunt Josephson junction stack can synchronize the other stack of IJJs, and vice versa. In this case one can obtain much stronger radiation power since both stacks are synchronized and contribute to the total radiation power. We show that the two stacks are synchronized along the whole IV curve. When the number of junctions in different stacks is the same, there is a nonzero phase shift between the plasma oscillation in the two stacks. When the number of junctions is different, the phase shift can be arbitrary. The phase shift can be tuned continuously by applying a magnetic field when junctions are connected by superconducting wires. 

\section{Model}
We consider two stacks of short IJJs biased by a dc current as sketched in Fig. \ref{f1}. We neglect the effect of radiation in the initial stage after switching on the bias current. Thus the superconducting phase is uniform in the lateral direction when the fluctuations are neglected. We then simplify the stack into one dimensional array of point-like Josephson junctions. The weak capacitive coupling between neighboring junctions is also neglected in the following calculations. \cite{Koyama96} The Josephson junction is modeled as a nonlinear inductor, shunted by a resistor and a capacitor as shown in Fig. \ref{f1} (b). It is known that even in the presence of small variation of parameters, the junctions can still be synchronized \cite{Bulaevskii07}. For simplicity we assume that the parameters for different junctions are identical. The equation of motion for the gauge invariant phase difference $\phi_l$, $\theta_l$ in the $l$-th junction can be written as
\begin{equation}\label{eq1}
\ddot{\phi }_l+\beta  \dot{\phi }_l+\sin  \phi _l-(I_{\mathrm{ext}}-I_{\theta })=0,
\end{equation}
\begin{equation}\label{eq2}
\ddot{\theta }_l+\beta  \dot{\theta }_l+\sin  \theta _l-I_{\theta }=0,
\end{equation}
\begin{equation}\label{eq3}
\sum _j^{N_{\phi }} \dot{\phi }_j=\sum _j^{N_{\theta }} \dot{\theta }_j,
\end{equation}
where the dimensionless units have been used: current is in units of the Josephson critical current $I_c$; frequency is in unit of the Josephson plasma frequency $\omega_p=\sqrt{2 e I_c/(\hbar C_J)}$ with $C_J$ the junction capacitance;  voltage is in units of $\hbar\omega_p/(2e)$.  $I_{\phi}$ and $I_{\theta}$ are the current, $\phi_l$ and $\theta_l$ are the phase difference at the left and right stack respectively. Here $\beta$ is the dimensionless conductivity $\beta\equiv {\sqrt{\hbar  }}/({ R_J \sqrt{2e I_c C_J}})$ with $R_J$ being the junction resistance. For BSCCO,  the phase dynamics in IJJs is underdamped, $\beta\ll 1$, which manifests as hysteretic IV curves.

\begin{figure}[b]
\psfig{figure=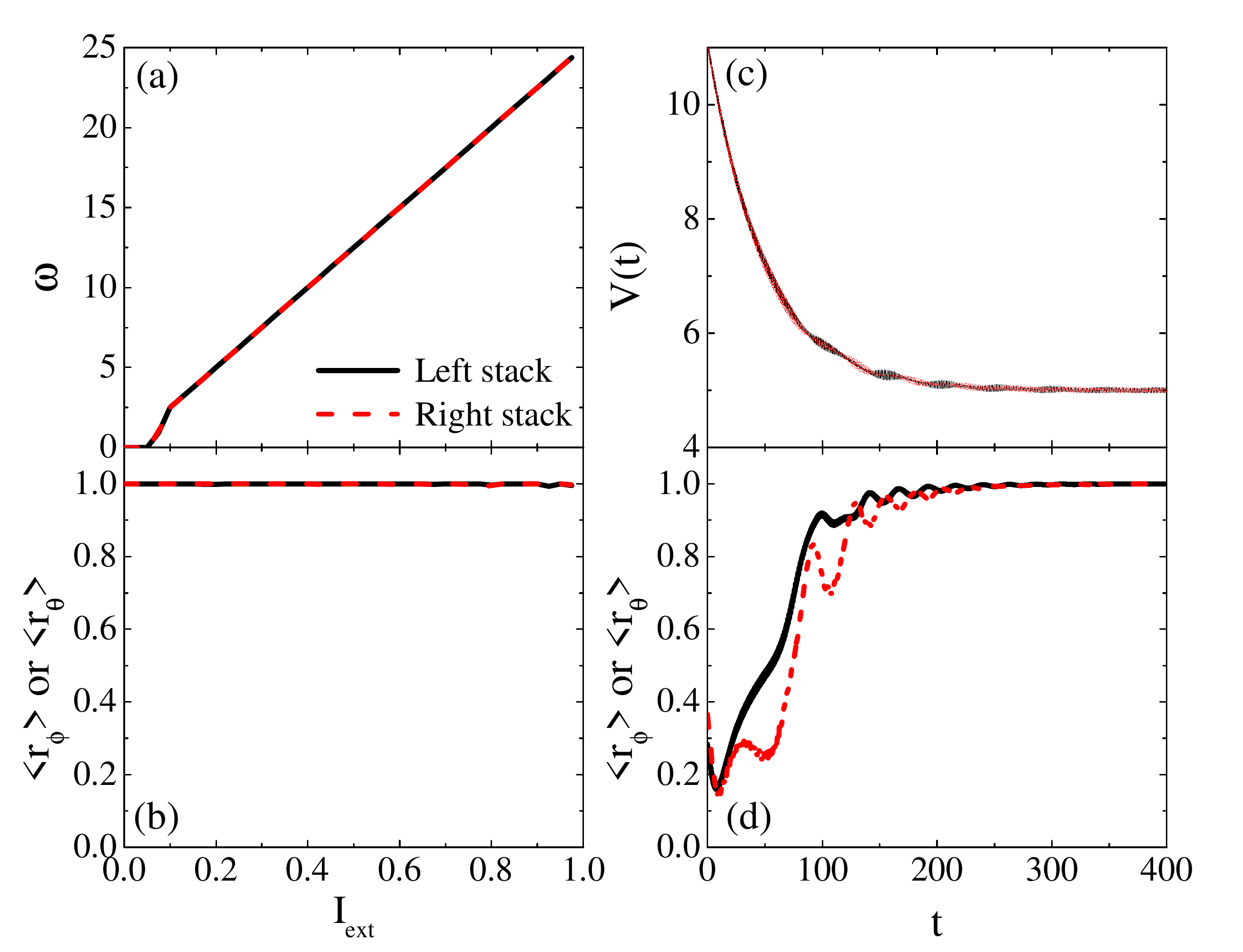,width=\columnwidth}
\caption{\label{f2}(color online)(a) IV curve of the left and right stacks. $\omega$ is the voltage per junction. (b) Amplitude of the order parameter versus the bias current. The stacks are completely synchronization for all currents. (c) Voltage per junction (d) amplitude of the order parameter as a function of time starting from a completely random state. The stacks are synchronized after $100/\omega_p$. Here $\beta=0.02$, $N_\phi=N_\theta=20$. $I_{\mathrm{ext}}=0.2$ for (c) and (d). We have also checked for $N_\phi=N_\theta=400$ and the results are qualitatively the same.}
\end{figure}

\subsection{The same number of junctions in the two stacks}

First we study the case that number of junctions in both stacks is the same $N_\phi=N_\theta$. When the junctions are fully synchronized in the stacks, i.e. $\phi_l=\phi_0$ and $\theta_l=\theta_0$, the solution in the voltage state with $\omega\gg 1$ can be written as
\begin{equation}\label{eq4}
\phi _0=\omega t +\mathrm{Re}[A_{\phi } e^  {i \omega t}]\ \ \ \ \mathrm{and}\ \ \ \  \theta_0=\omega t+\gamma+\mathrm{Re}[A_{\theta } e^{i (\gamma+ \omega t )}],
\end{equation}
where we have accounted for the possible phase shift $\gamma$ between the left and right stack. Substituting Eq. \eqref{eq4} into Eqs. \eqref{eq1}, \eqref{eq2} and \eqref{eq3} and comparing each component with the same frequency, we obtain
\begin{equation}\label{eq5}
{A_\phi } = \frac{{i[1 + \exp ({i}\gamma )]}}{{2( - {\omega ^2} + {i}\beta \omega )}} \ \ \ \ \mathrm{and}\ \ \ \ {A_\theta } = \frac{{i[1 + \exp (-{i}\gamma )]}}{{2( - {\omega ^2} + {i}\beta \omega )}}. 
\end{equation}
Note the voltage at both stack is equal at any time according to Eq. \eqref{eq3} , i.e. $\dot{\phi}_0(t)=\dot{\theta}_0(t)$. The corresponding IV characteristics is
\begin{equation}\label{eq6}
I_{\mathrm{ext}} = 2\beta \omega  + \frac{{\beta \omega (1 + \cos\gamma )}}{{2({\omega ^4} + {\beta ^2}{\omega ^2})}}.
\end{equation}
We then study the linear stability of the uniform solution by adding a small perturbation to the uniform solution $\phi_l =\phi _0+\tilde{\phi }_l$, $\theta_l =\theta_0+\tilde{\theta }_l$. The system has the permutation symmetry, i.e. does not depend on the order of the Josephson junction in the stack, and we introduce $\Delta_{\phi, l}=\phi_{l+1}-\phi_l$. The equation for $\Delta_{\phi, l}$ is
\begin{equation}\label{eq7}
\ddot{ \Delta}_{\phi,l}  + \beta {\dot \Delta_{\phi,l} } + \cos(\phi _0){\Delta_{\phi,l} } = 0,
\end{equation}
and similarly for $\Delta_{\theta, l}$. The uniform solution is stable only when $\Delta_{\phi, l}$ and $\Delta_{\theta, l}$ decay with time. The perturbation $\Delta_{\phi, l}$ couples nonlinearly with the oscillation $\cos(\phi_0)$ and frequency harmonics are induced. The solution to Eq. \eqref{eq7} can be written as
\begin{equation}\label{eq8}
\Delta _{\phi,l } = ({a_0} + {a_ + }{{{e}}^{{{i}}\omega t}} + {a_ - }{{{e}}^{ - {{i}}\omega t}}){{{e}}^{ - {{i}}\Omega t}},
\end{equation}
with $\Omega\ll 1$. The uniform solution is stable when and only when $\mathrm{Im}[\Omega]<0$ for all solutions $\Omega$. We have neglected higher frequency components $p \omega-\Omega$ with integer $|p|>1$ because they are small, $a_p\sim 1/(p^2 \omega^2)$, when $\omega\gg 1$. Substituting Eq. \eqref{eq8} into Eq. \eqref{eq7} and comparing each frequency component, we have 
\begin{equation}\label{eq9}
 - {\omega ^2}{a_ {\pm} } \pm {{i}}\omega \beta {a_{\pm}} + \frac{1}{2}{a_0} = 0,
\end{equation}
\begin{equation}\label{eq10}
- {\Omega ^2}{a_0} - {{i}}\beta \Omega {a_0} + \mathrm{Re}\left[\frac{{{A_\phi }}}{{2{{i}}}}\right]{a_0} + \frac{1}{2}({a_ + } + {a_ - }) = 0.
\end{equation}
From Eqs. \eqref{eq9} and \eqref{eq10} we obtain the equation for $\Omega$ which determines the spectrum of the perturbation $\Delta_{\phi, l}$
\begin{equation}\label{eq11}
{\Omega ^2} + {{i}}\beta \Omega  = \frac{{[1 - \cos \gamma ]}}{{4{\omega ^2}}}.
\end{equation}
The spectrum for the perturbation $\Delta_{\theta, l}$ is the same. The uniform solution therefore is stable for any $\gamma\neq 0$.

To go beyond the above linear stability analysis, we also solve Eqs. \eqref{eq1}, \eqref{eq2} and \eqref{eq3} numerically with a random initial state. The degree of synchronization is captured by the order parameter\cite{Acebron05}
\begin{equation}\label{eq12}
\Psi_\phi=r_\phi(t)\exp[i\varphi_\phi(t)]=\frac{1}{N_{\phi}}\sum_j^{N_{\phi}} \exp(i\phi_j).
\end{equation}
Here $r_{\phi}$ is positively defined. We compute the average of $r_{\phi}(t)$
\begin{equation}\label{eq13}
\langle r_{\phi} \rangle=\frac{1}{t_f}\int_{0}^{t_f} dt\ r_{\phi}(t).
\end{equation}
and take $t_f\rightarrow+\infty$. In the fully synchronized state $\langle r_{\phi} \rangle=1$. The IV characteristic presented in Fig. \ref{f2}(a) is typical for underdamped junctions: starting from a superconducting state at $I_{\mathrm{ext}}=0$, the system switches into the voltage state at $I_\theta>1$ and $I_\phi>1$. The system remains in the voltage state even when $I_\phi$ or $I_\theta$ are reduced below the critical current. The IV curve at high voltage is linear. The system retraps into zero-voltage state once the current is below a threshold current or retrapping current. As shown in Fig. \ref{f2} (b) $\langle r_\phi\rangle=1$ and $\langle r_\theta\rangle=1$ indicating that the junction are fully synchronized in the voltage state, which is consistent with the analytical calculations. The time dependence of voltage, $\langle r_{\phi}\rangle$ and $\langle r_{\theta}\rangle$ is depicted in Fig. \ref{f2} (c) and (d). The stacks are fully synchronized after $100/\omega_p$ at $I_{\mathrm{ext}}=0.2$. The synchronization time decreases with current.

\begin{figure}[t]
\psfig{figure=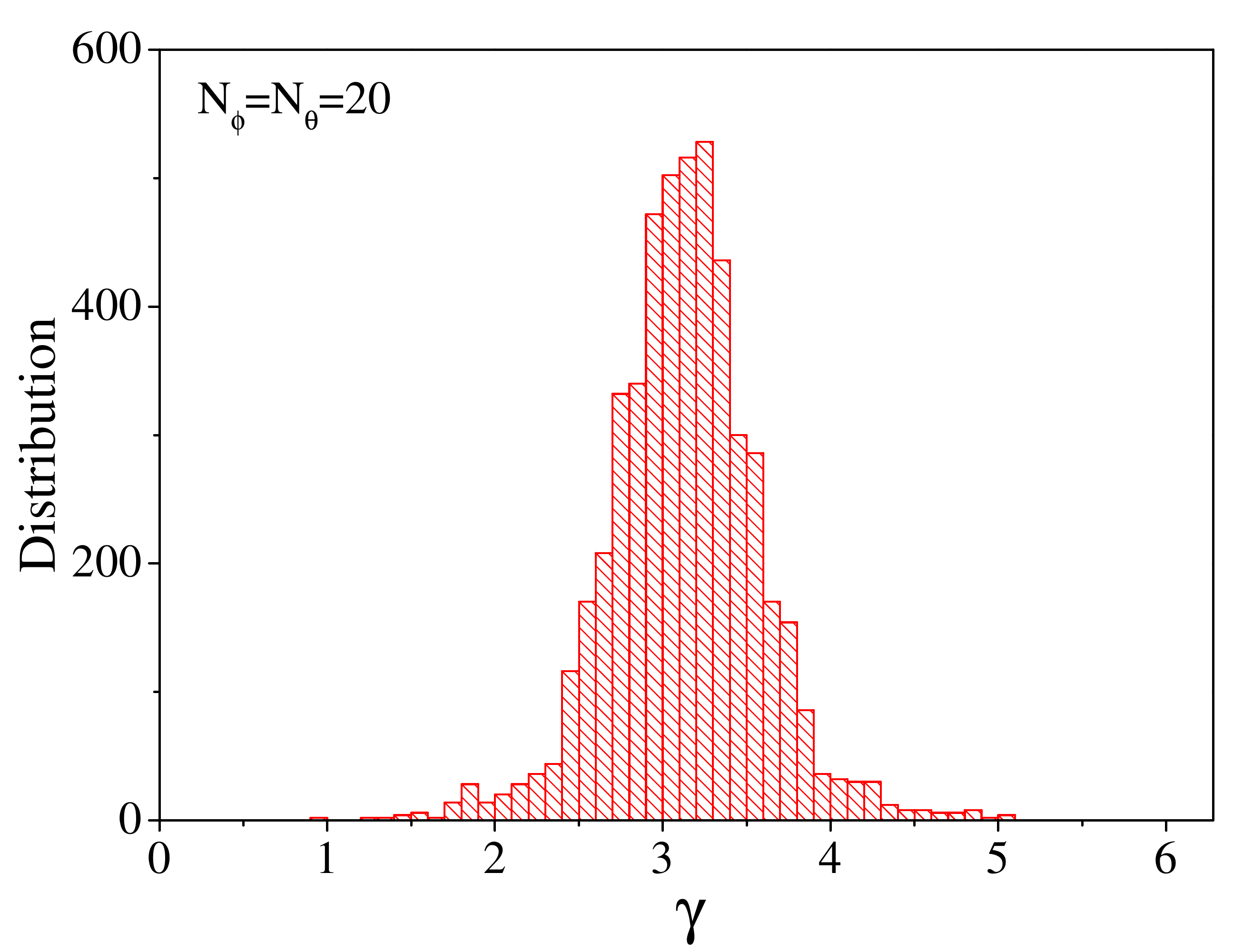,width=\columnwidth}
\caption{\label{f3}(color online) Distribution of the phase shift between the two stacks obtained by repeating the numerical simulations of Eqs. \eqref{eq1}, \eqref{eq2} and \eqref{eq3} with different random initial states. Here $\beta=0.02$, $N_\phi=N_\theta=20$ and $I_{\mathrm{ext}}=0.2$.}
\end{figure}

In the synchronized state, there is an undetermined phase shift between the left and right stack suggested from the above linear stability analysis. The phase difference $\gamma$ can be determined from the order parameter $\Psi_\phi$ and $\Psi_\theta$, $\gamma=\mathrm{arg}(\Psi_\phi\Psi_\theta^*)$. We calculate the distribution of $\gamma$ by repeating the simulation $5000$ times with different random initial states, and the results are shown in Fig. \ref{f3}. The distribution of $\gamma$ has a peak around $\pi$ because $\gamma=\pi$ is the most stable case according to Eq. \eqref{eq11}. Contrarily, the distribution around $\gamma=0$ is zero because the synchronized state is neutrally stable for $\gamma=0$. The dc current contribution due to the plasma oscillation [the second term at the right hand side in Eq. \eqref{eq6}], which is proportional to $1+\cos\gamma$, is small, and the IV curve becomes ohmic $I_{\mathrm{ext}}\approx 2\beta\omega$.

\begin{figure}[b]
\psfig{figure=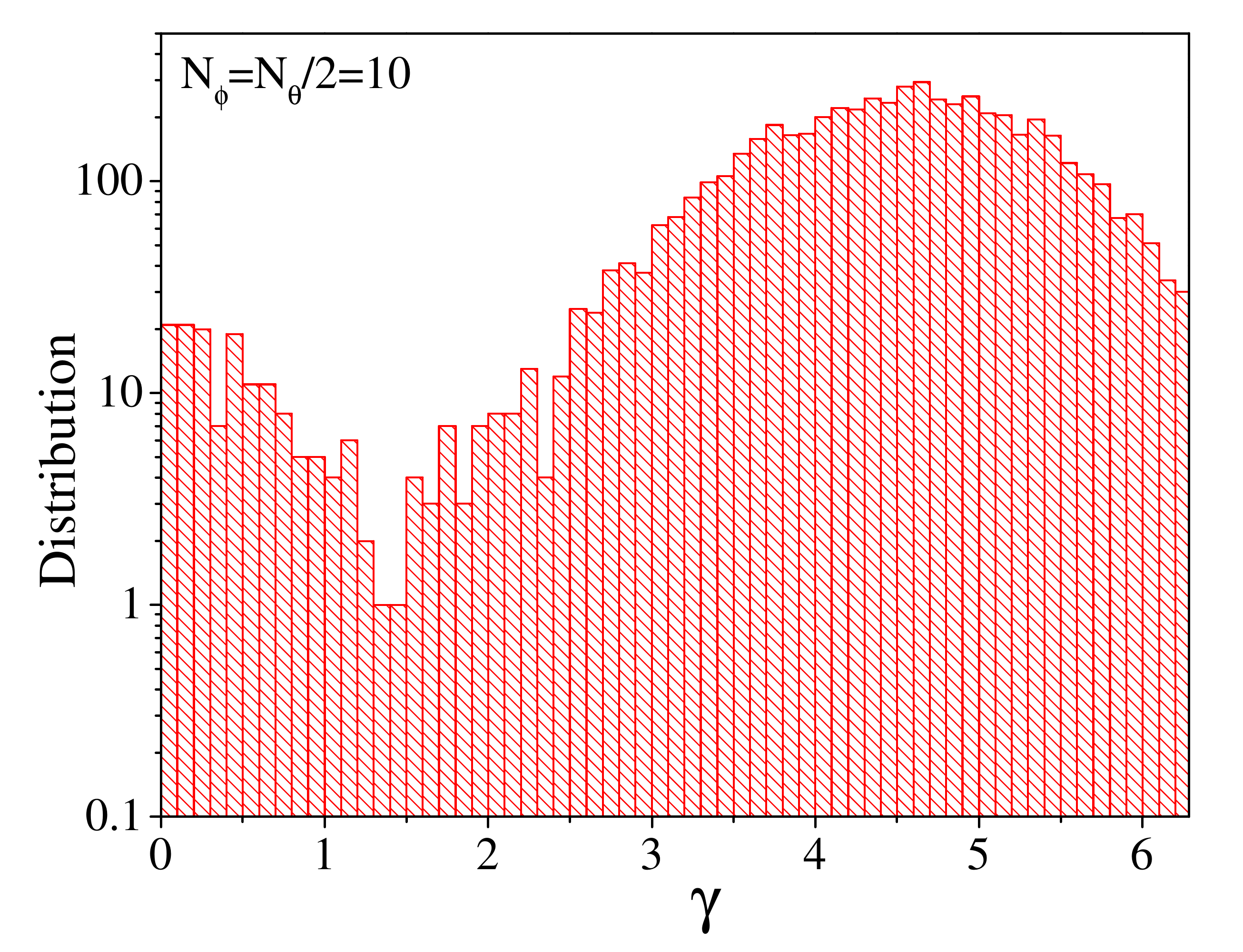,width=\columnwidth}
\caption{\label{f4}(color online) The same as Fig. \ref{f3} but with $N_\phi=N_\theta/2=10$.}
\end{figure}

\subsection{Number of junctions in the two stack is different}

We proceed to investigate the situation when $N_\phi\neq N_\theta$. In this case, the voltage per junction at the two stacks are different, $N_\phi\omega_\phi= N_\theta\omega_\theta$. As the total voltage at the two stacks is equal at any time, there are plasma oscillation both at frequencies $\omega_\phi$ and $\omega_\theta$ in the two stacks in the synchronized state. The solution $\theta_0$ and $\phi_0$ in the region $\omega_\theta\gg 1$ and $\omega_\phi\gg 1$ can be written as
\begin{equation}\label{eq14}
{\phi _0} = {\omega _\phi}t + \mathrm{Re}[ {A_{\phi 1}}\exp ({{i}}{\omega _\phi}t) ] + \mathrm{Re}[{A_{\phi 2}}\exp ({{i}}{\omega _\theta}t)],
\end{equation}  
\begin{equation}\label{eq15}
{\theta _0} = {\omega _\theta}t + \gamma  + \mathrm{Re}[ {A_{\theta 1}}\exp({{i}}{\omega _\phi}t) ] + \mathrm{Re}[{A_{\theta 2}}\exp ({{i}}{\omega _\theta}t)].
\end{equation}  
Again we have taken the possible phase shift $\gamma$ between the two stacks into account. The amplitude of the plasma oscillation is
\[
{A_{\phi 1}} = \frac{{{{i}}}}{{(1 + \frac{{{N_\phi }}}{{{N_\theta }}})( - \omega _\phi^2 + {{i}}\beta {\omega _\phi})}},\ \ \ \ {A_{\phi 2}} = \frac{{{{i}}\exp ({{i}}\gamma )}}{{(1 + \frac{{{N_\phi }}}{{{N_\theta }}})( - \omega _\theta^2 + {{i}}\beta {\omega _\theta})}}.
\]
\[
{A_{\theta 1}} = \frac{{{{i}}}}{{(1 + \frac{{{N_\theta }}}{{{N_\phi }}})( - \omega _\phi^2 + {{i}}\beta {\omega _\phi})}},\ \ \ \ {A_{\theta 2}} = \frac{{{{i}}\exp ({{i}}\gamma )}}{{(1 + \frac{{{N_\theta }}}{{{N_\phi }}})( - \omega _\theta^2 + {{i}}\beta {\omega _\theta})}}.
\]
The IV characteristic is $I_{\mathrm{ext}}=\beta ({\omega _\phi} + \omega_\theta )+I_p$ with the plasma oscillation contribution $I_p$
\begin{equation}\label{eq16}
I_{p}=\frac{{\beta {\omega _\phi}}}{{2(1 + \frac{{{N_\phi }}}{{{N_\theta }}})(\omega _\phi^4 + {\beta ^2}\omega _\phi^2)}} + \frac{{\beta {\omega _2}}}{{2(1 + \frac{{{N_\theta }}}{{{N_\phi }}})(\omega _\theta^4 + {\beta ^2}\omega _\theta^2)}}.
\end{equation}

We then perform the similar stability analysis for the uniform solution as done for the $N_\phi=N_\theta$ case. The perturbation with frequency $\Omega\ll 1$ and amplitude $a_0$ will induce higher harmonics $p\omega_\phi-\Omega$ with amplitude ${(p{\omega _\phi})^{ - 2}}{a_0}$ and also the frequency component $\omega_\theta-\omega_\phi-\Omega$ with amplitude ${({\omega _\theta} - {\omega _\phi} - \Omega )^{ - 2}}{A_{\phi 2}}{a_0}$. We assume that $\omega_\theta$ and $\omega_\phi$ are well separated $|\omega_\phi-\omega_\theta|\gg 1$ or $|1 - {N_\phi }/{N_\theta }|{\omega _\phi}\gg1$. The frequency component $\omega_\theta-\omega_\phi-\Omega$ has amplitude of the order of ${A_{\phi 2}}/{({\omega _\phi} - {\omega _\theta})^2}$, which is small compared to $1/\omega_\phi^2$. We may neglect the frequency component $\omega_\theta-\omega_\phi-\Omega$, and we obtain the spectrum for the perturbations with respect to $\phi_0$
\begin{equation}\label{eq17}
{\Omega ^2} + {{i}}\beta \Omega  = \frac{{{N_\phi }}}{{2({N_\theta } + {N_\phi })\omega _\phi^2}}.
\end{equation} 
The spectrum for the perturbations with respect to $\theta_0$ is obtained by replacing the expression in the right hand side of Eq. \eqref{eq17} by $\frac{{{N_\theta }}}{{2({N_\theta } + {N_\phi })\omega _\theta^2}}$. The perturbations always decay with time because $\mathrm{Im}[\Omega]<0$, thus the uniform solution is stable.  Note here $\Omega$ is independent of $\gamma$, which is different from the case with $N_\phi=N_\theta$. We have also calculated Eqs. \eqref{eq1}, \eqref{eq2} and \eqref{eq3} numerical with $N_\phi=N_\theta/2=10$ starting from a random initial state and the results are qualitatively similar to those in Fig. \ref{f2} for $N_\phi=N_\theta$. However for the phase difference between the two stacks in the synchronized state as shown in Fig. \ref{f4}, we found that $\gamma$ can be any value in contrast to the case $N_\phi=N_\theta$ where only $\gamma\neq0$ is allowed. Note that the distribution of $\gamma$ is not uniform which indicates that the basin of attraction for certain $\gamma$ occupies larger volume in the phase space.

\section{Discussions and conclusions}
Here we propose to achieve mutual synchronization by shunting two stacks of short intrinsic Josephson junctions in parallel. The junctions are fully synchronized after several hundreds of $1/\omega_p$ starting from a completely disordered state. In the synchronized state, the strong radiation fields also tend to stabilize the uniform oscillating state. \cite{Bulaevskii07} When the number of junctions between the two stacks is the same, there is a nonzero phase shift between the two stacks. According to Fig. \ref{f3}, the most probable phase shift is $\gamma\approx \pi$. This out-of-phase oscillation in the two stacks is harmful for obtaining strong THz radiation. One may introduce a mirror to separate the radiation fields from the two stacks. When the number of the junctions between the two stacks is different, the phase shift can be arbitrary. In both cases, one may tune the phase shift by using superconducting wires to connect these junctions and applying an external magnetic field in the loop closed by the two stacks. The phase shift is determined by $\sum_{l=1}^{N_\phi}\phi_l-\sum_{l=1}^{N_\theta}\theta_l=(p+\Phi/\Phi_0)2\pi$. Here $\Phi$ is the flux in the loop closed by the two stacks and $\Phi_0=hc/(2e)$ is the quantum flux. When $\Phi$ varies continuously, the phase shift also changes correspondingly which affects the far-field radiation pattern. The IV characteristics for BSCCO is extremely hysteretic, and it may be difficult to control the number of junctions in the voltage state in both stacks. Therefore the case with $N_\phi\neq N_\theta$ is more relevant to experiments. 

We then compare the synchronization of multiple mesas through the substrate with the present proposal. Recently multiple mesas were fabricated atop of BSCCO single crystal, where much strong radiation power was achieved by synchronizing these mesas. \cite{Orita2010,Benseman12} In Ref. \onlinecite{szlin13} it is demonstrated that the mesas can be synchronized via the plasma oscillation in the basal crystal. The plasma in mesas propagate into the basal crystal and then establish mutual interaction among the plasma oscillation in different mesas. This coupling is strongest when the width and position of the mesa are commensurate with the wave structure of plasma oscillation in the basal crystal. Meanwhile the synchronization is achieved only when the bias voltage/current is tuned such that the cavity resonance condition for each mesa is satisfied. The present design using two stacks of junctions is much simpler and easier to achieve synchronization. The two stacks are fully synchronized in the whole IV curve. Moreover the phase difference of the radiation fields from the two stacks can be controlled with an external magnetic field.  

Finally we discuss the effect of thermal fluctuations on the synchronized state. Generally speaking the thermal noise destroys the synchronized at a critical temperature $T_m$, i.e. the melting of the synchronized state. Although there is an arbitrary phase shift between the two stacks in the case with $N_\phi\neq N_\theta$, the synchronized state is robust against weak fluctuations, because the effective dimensionality of the system is infinite or mean-field type. The junction is coupled with the current through it, which depends on the phase dynamics of the rest junctions in the circuit. In this way, the junction is coupled to rest of junctions. This is crucial for synchronization since the lower critical dimension for the phase synchronize is four. \cite{Hong05} The melting transition in the case of a capacitor shunt was studied \cite{szlin11} and the melting is a second order phase transition $\langle  r_\phi \rangle\sim (T_m-T)^{\eta}$ with the exponent $\eta=1/2$, consistent with the mean-field expectation value. We expect a similar mean-field type melting transition when the temperature is increased in the present system. 

It is well known that the junctions can also be synchronized for non-identical junctions. \cite{Hadley88} We have performed simulation for junctions with different critical currents, i.e. the critical current increases linearly from $0.9I_c$ to $1.1 I_c$ for $N_\phi=N_\theta=20$ at both stacks. The results are qualitatively the same as those in Fig. \ref{f2} (d).

To summarize, we have proposed to synchronize a stack of intrinsic Josephson junctions by shunting another stack of junctions in parallel. The junctions are fully synchronized in the whole IV curve. When the number of the junctions in different stacks is identical, there is a nonzero phase shift between the plasma oscillation in different stacks. When the number of junctions is different, the phase shift can be arbitrary. This phase shift can be tuned continuously by applying an external magnetic field when all junctions are connected by superconducting wires. 

\section{Acknowledgments}
We thank L. N. Bulaevskii for useful discussions. This work was carried out under the auspices of the NNSA of the U.S. DOE at LANL under Award No. DE-AC52-06NA25396.


\begin{thebibliography}{47}%
\makeatletter
\providecommand \@ifxundefined [1]{%
 \@ifx{#1\undefined}
}%
\providecommand \@ifnum [1]{%
 \ifnum #1\expandafter \@firstoftwo
 \else \expandafter \@secondoftwo
 \fi
}%
\providecommand \@ifx [1]{%
 \ifx #1\expandafter \@firstoftwo
 \else \expandafter \@secondoftwo
 \fi
}%
\providecommand \natexlab [1]{#1}%
\providecommand \enquote  [1]{``#1''}%
\providecommand \bibnamefont  [1]{#1}%
\providecommand \bibfnamefont [1]{#1}%
\providecommand \citenamefont [1]{#1}%
\providecommand \href@noop [0]{\@secondoftwo}%
\providecommand \href [0]{\begingroup \@sanitize@url \@href}%
\providecommand \@href[1]{\@@startlink{#1}\@@href}%
\providecommand \@@href[1]{\endgroup#1\@@endlink}%
\providecommand \@sanitize@url [0]{\catcode `\\12\catcode `\$12\catcode
  `\&12\catcode `\#12\catcode `\^12\catcode `\_12\catcode `\%12\relax}%
\providecommand \@@startlink[1]{}%
\providecommand \@@endlink[0]{}%
\providecommand \url  [0]{\begingroup\@sanitize@url \@url }%
\providecommand \@url [1]{\endgroup\@href {#1}{\urlprefix }}%
\providecommand \urlprefix  [0]{URL }%
\providecommand \Eprint [0]{\href }%
\providecommand \doibase [0]{http://dx.doi.org/}%
\providecommand \selectlanguage [0]{\@gobble}%
\providecommand \bibinfo  [0]{\@secondoftwo}%
\providecommand \bibfield  [0]{\@secondoftwo}%
\providecommand \translation [1]{[#1]}%
\providecommand \BibitemOpen [0]{}%
\providecommand \bibitemStop [0]{}%
\providecommand \bibitemNoStop [0]{.\EOS\space}%
\providecommand \EOS [0]{\spacefactor3000\relax}%
\providecommand \BibitemShut  [1]{\csname bibitem#1\endcsname}%
\let\auto@bib@innerbib\@empty
\bibitem [{\citenamefont {Yanson}\ \emph {et~al.}(1965)\citenamefont {Yanson},
  \citenamefont {Svistunov},\ and\ \citenamefont {Dmitrenko}}]{Yanson65}%
  \BibitemOpen
  \bibfield  {author} {\bibinfo {author} {\bibfnamefont {I.~K.}\ \bibnamefont
  {Yanson}}, \bibinfo {author} {\bibfnamefont {V.~M.}\ \bibnamefont
  {Svistunov}}, \ and\ \bibinfo {author} {\bibfnamefont {I.~M.}\ \bibnamefont
  {Dmitrenko}},\ }\href@noop {} {\bibfield  {journal} {\bibinfo  {journal} {Zh.
  Eksp. Teor. Fiz.}\ }\textbf {\bibinfo {volume} {48}},\ \bibinfo {pages} {976}
  (\bibinfo {year} {1965})}\BibitemShut {NoStop}%
\bibitem [{\citenamefont {Dayem}\ and\ \citenamefont {Grimes}(1966)}]{Dayem66}%
  \BibitemOpen
  \bibfield  {author} {\bibinfo {author} {\bibfnamefont {A.~H.}\ \bibnamefont
  {Dayem}}\ and\ \bibinfo {author} {\bibfnamefont {C.~C.}\ \bibnamefont
  {Grimes}},\ }\href@noop {} {\bibfield  {journal} {\bibinfo  {journal} {Appl.
  Phys. Lett.}\ }\textbf {\bibinfo {volume} {9}},\ \bibinfo {pages} {47}
  (\bibinfo {year} {1966})}\BibitemShut {NoStop}%
\bibitem [{\citenamefont {Zimmerma.Je}\ \emph {et~al.}(1966)\citenamefont
  {Zimmerma.Je}, \citenamefont {Cowen},\ and\ \citenamefont
  {Silver}}]{Zimmerma66}%
  \BibitemOpen
  \bibfield  {author} {\bibinfo {author} {\bibnamefont {Zimmerma.Je}}, \bibinfo
  {author} {\bibfnamefont {J.~A.}\ \bibnamefont {Cowen}}, \ and\ \bibinfo
  {author} {\bibfnamefont {A.~H.}\ \bibnamefont {Silver}},\ }\href@noop {}
  {\bibfield  {journal} {\bibinfo  {journal} {Appl. Phys. Lett.}\ ,\ \bibinfo
  {pages} {353}} (\bibinfo {year} {1966})}\BibitemShut {NoStop}%
\bibitem [{\citenamefont {Bulaevskii}\ and\ \citenamefont
  {Koshelev}(2006{\natexlab{a}})}]{Bulaevskii06PRL}%
  \BibitemOpen
  \bibfield  {author} {\bibinfo {author} {\bibfnamefont {L.~N.}\ \bibnamefont
  {Bulaevskii}}\ and\ \bibinfo {author} {\bibfnamefont {A.~E.}\ \bibnamefont
  {Koshelev}},\ }\href@noop {} {\bibfield  {journal} {\bibinfo  {journal}
  {Phys. Rev. Lett.}\ }\textbf {\bibinfo {volume} {97}},\ \bibinfo {pages}
  {267001} (\bibinfo {year} {2006}{\natexlab{a}})}\BibitemShut {NoStop}%
\bibitem [{\citenamefont {Finnegan}\ and\ \citenamefont
  {Wahlsten}(1972)}]{Finnegan72}%
  \BibitemOpen
  \bibfield  {author} {\bibinfo {author} {\bibfnamefont {T.~F.}\ \bibnamefont
  {Finnegan}}\ and\ \bibinfo {author} {\bibfnamefont {S.}~\bibnamefont
  {Wahlsten}},\ }\href@noop {} {\bibfield  {journal} {\bibinfo  {journal}
  {Appl. Phys. Lett.}\ }\textbf {\bibinfo {volume} {21}},\ \bibinfo {pages}
  {541} (\bibinfo {year} {1972})}\BibitemShut {NoStop}%
\bibitem [{\citenamefont {Jain}\ \emph {et~al.}(1984)\citenamefont {Jain},
  \citenamefont {Likharev}, \citenamefont {Lukens},\ and\ \citenamefont
  {Sauvageau}}]{Jain84}%
  \BibitemOpen
  \bibfield  {author} {\bibinfo {author} {\bibfnamefont {A.~K.}\ \bibnamefont
  {Jain}}, \bibinfo {author} {\bibfnamefont {K.~K.}\ \bibnamefont {Likharev}},
  \bibinfo {author} {\bibfnamefont {J.~E.}\ \bibnamefont {Lukens}}, \ and\
  \bibinfo {author} {\bibfnamefont {J.~E.}\ \bibnamefont {Sauvageau}},\
  }\href@noop {} {\bibfield  {journal} {\bibinfo  {journal} {Phys. Rep.}\
  }\textbf {\bibinfo {volume} {109}},\ \bibinfo {pages} {309} (\bibinfo {year}
  {1984})}\BibitemShut {NoStop}%
\bibitem [{\citenamefont {Darula}\ \emph {et~al.}(1999)\citenamefont {Darula},
  \citenamefont {Doderer},\ and\ \citenamefont {Beuven}}]{Durala99}%
  \BibitemOpen
  \bibfield  {author} {\bibinfo {author} {\bibfnamefont {M.}~\bibnamefont
  {Darula}}, \bibinfo {author} {\bibfnamefont {T.}~\bibnamefont {Doderer}}, \
  and\ \bibinfo {author} {\bibfnamefont {S.}~\bibnamefont {Beuven}},\
  }\href@noop {} {\bibfield  {journal} {\bibinfo  {journal} {Supercond. Sci.
  Technol.}\ }\textbf {\bibinfo {volume} {12}},\ \bibinfo {pages} {R1}
  (\bibinfo {year} {1999})}\BibitemShut {NoStop}%
\bibitem [{\citenamefont {Barbara}\ \emph {et~al.}(1999)\citenamefont
  {Barbara}, \citenamefont {Cawthorne}, \citenamefont {Shitov},\ and\
  \citenamefont {Lobb}}]{Barbara99}%
  \BibitemOpen
  \bibfield  {author} {\bibinfo {author} {\bibfnamefont {P.}~\bibnamefont
  {Barbara}}, \bibinfo {author} {\bibfnamefont {A.~B.}\ \bibnamefont
  {Cawthorne}}, \bibinfo {author} {\bibfnamefont {S.~V.}\ \bibnamefont
  {Shitov}}, \ and\ \bibinfo {author} {\bibfnamefont {C.~J.}\ \bibnamefont
  {Lobb}},\ }\href@noop {} {\bibfield  {journal} {\bibinfo  {journal} {Phys.
  Rev. Lett.}\ }\textbf {\bibinfo {volume} {82}},\ \bibinfo {pages} {1963}
  (\bibinfo {year} {1999})}\BibitemShut {NoStop}%
\bibitem [{\citenamefont {Song}\ \emph {et~al.}(2009)\citenamefont {Song},
  \citenamefont {M¨¹ller}, \citenamefont {Behr},\ and\ \citenamefont
  {Klushin}}]{Song09}%
  \BibitemOpen
  \bibfield  {author} {\bibinfo {author} {\bibfnamefont {F.}~\bibnamefont
  {Song}}, \bibinfo {author} {\bibfnamefont {F.}~\bibnamefont {M¨¹ller}},
  \bibinfo {author} {\bibfnamefont {R.}~\bibnamefont {Behr}}, \ and\ \bibinfo
  {author} {\bibfnamefont {A.~M.}\ \bibnamefont {Klushin}},\ }\href@noop {}
  {\bibfield  {journal} {\bibinfo  {journal} {Appl. Phys. Lett.}\ }\textbf
  {\bibinfo {volume} {95}},\ \bibinfo {pages} {172501} (\bibinfo {year}
  {2009})}\BibitemShut {NoStop}%
\bibitem [{\citenamefont {Kleiner}\ \emph {et~al.}(1992)\citenamefont
  {Kleiner}, \citenamefont {Steinmeyer}, \citenamefont {Kunkel},\ and\
  \citenamefont {M\"{u}ller}}]{Kleiner92}%
  \BibitemOpen
  \bibfield  {author} {\bibinfo {author} {\bibfnamefont {R.}~\bibnamefont
  {Kleiner}}, \bibinfo {author} {\bibfnamefont {F.}~\bibnamefont {Steinmeyer}},
  \bibinfo {author} {\bibfnamefont {G.}~\bibnamefont {Kunkel}}, \ and\ \bibinfo
  {author} {\bibfnamefont {P.}~\bibnamefont {M\"{u}ller}},\ }\href@noop {}
  {\bibfield  {journal} {\bibinfo  {journal} {Phys. Rev. Lett.}\ }\textbf
  {\bibinfo {volume} {68}},\ \bibinfo {pages} {2394} (\bibinfo {year}
  {1992})}\BibitemShut {NoStop}%
\bibitem [{\citenamefont {Iguchi}\ \emph {et~al.}(2000)\citenamefont {Iguchi},
  \citenamefont {Lee}, \citenamefont {Kume}, \citenamefont {Ishibashi},\ and\
  \citenamefont {Sato}}]{Iguchi00b}%
  \BibitemOpen
  \bibfield  {author} {\bibinfo {author} {\bibfnamefont {I.}~\bibnamefont
  {Iguchi}}, \bibinfo {author} {\bibfnamefont {K.}~\bibnamefont {Lee}},
  \bibinfo {author} {\bibfnamefont {E.}~\bibnamefont {Kume}}, \bibinfo {author}
  {\bibfnamefont {T.}~\bibnamefont {Ishibashi}}, \ and\ \bibinfo {author}
  {\bibfnamefont {K.}~\bibnamefont {Sato}},\ }\href@noop {} {\bibfield
  {journal} {\bibinfo  {journal} {Phys. Rev. B}\ }\textbf {\bibinfo {volume}
  {61}},\ \bibinfo {pages} {689} (\bibinfo {year} {2000})}\BibitemShut
  {NoStop}%
\bibitem [{\citenamefont {Batov}\ \emph {et~al.}(2006)\citenamefont {Batov},
  \citenamefont {Jin}, \citenamefont {Shitov}, \citenamefont {Koval},
  \citenamefont {M\"{u}ller},\ and\ \citenamefont {Ustinov}}]{Batov06}%
  \BibitemOpen
  \bibfield  {author} {\bibinfo {author} {\bibfnamefont {I.~E.}\ \bibnamefont
  {Batov}}, \bibinfo {author} {\bibfnamefont {X.~Y.}\ \bibnamefont {Jin}},
  \bibinfo {author} {\bibfnamefont {S.~V.}\ \bibnamefont {Shitov}}, \bibinfo
  {author} {\bibfnamefont {Y.}~\bibnamefont {Koval}}, \bibinfo {author}
  {\bibfnamefont {P.}~\bibnamefont {M\"{u}ller}}, \ and\ \bibinfo {author}
  {\bibfnamefont {A.~V.}\ \bibnamefont {Ustinov}},\ }\href@noop {} {\bibfield
  {journal} {\bibinfo  {journal} {Appl. Phys. Lett.}\ }\textbf {\bibinfo
  {volume} {88}},\ \bibinfo {pages} {262504} (\bibinfo {year}
  {2006})}\BibitemShut {NoStop}%
\bibitem [{\citenamefont {Bae}\ \emph {et~al.}(2007)\citenamefont {Bae},
  \citenamefont {Lee},\ and\ \citenamefont {Choi}}]{Bae07}%
  \BibitemOpen
  \bibfield  {author} {\bibinfo {author} {\bibfnamefont {M.~H.}\ \bibnamefont
  {Bae}}, \bibinfo {author} {\bibfnamefont {H.~J.}\ \bibnamefont {Lee}}, \ and\
  \bibinfo {author} {\bibfnamefont {J.~H.}\ \bibnamefont {Choi}},\ }\href@noop
  {} {\bibfield  {journal} {\bibinfo  {journal} {Phys. Rev. Lett.}\ }\textbf
  {\bibinfo {volume} {98}},\ \bibinfo {pages} {027002} (\bibinfo {year}
  {2007})}\BibitemShut {NoStop}%
\bibitem [{\citenamefont {Benseman}\ \emph {et~al.}(2011)\citenamefont
  {Benseman}, \citenamefont {Koshelev}, \citenamefont {Gray}, \citenamefont
  {Kwok}, \citenamefont {Welp}, \citenamefont {Kadowaki}, \citenamefont
  {Tachiki},\ and\ \citenamefont {Yamamoto}}]{Benseman11}%
  \BibitemOpen
  \bibfield  {author} {\bibinfo {author} {\bibfnamefont {T.~M.}\ \bibnamefont
  {Benseman}}, \bibinfo {author} {\bibfnamefont {A.~E.}\ \bibnamefont
  {Koshelev}}, \bibinfo {author} {\bibfnamefont {K.~E.}\ \bibnamefont {Gray}},
  \bibinfo {author} {\bibfnamefont {W.-K.}\ \bibnamefont {Kwok}}, \bibinfo
  {author} {\bibfnamefont {U.}~\bibnamefont {Welp}}, \bibinfo {author}
  {\bibfnamefont {K.}~\bibnamefont {Kadowaki}}, \bibinfo {author}
  {\bibfnamefont {M.}~\bibnamefont {Tachiki}}, \ and\ \bibinfo {author}
  {\bibfnamefont {T.}~\bibnamefont {Yamamoto}},\ }\href@noop {} {\bibfield
  {journal} {\bibinfo  {journal} {Phys. Rev. B}\ }\textbf {\bibinfo {volume}
  {84}},\ \bibinfo {pages} {064523} (\bibinfo {year} {2011})}\BibitemShut
  {NoStop}%
\bibitem [{\citenamefont {Tachiki}\ \emph {et~al.}(1994)\citenamefont
  {Tachiki}, \citenamefont {Koyama},\ and\ \citenamefont
  {Takahashi}}]{Tachiki94}%
  \BibitemOpen
  \bibfield  {author} {\bibinfo {author} {\bibfnamefont {M.}~\bibnamefont
  {Tachiki}}, \bibinfo {author} {\bibfnamefont {T.}~\bibnamefont {Koyama}}, \
  and\ \bibinfo {author} {\bibfnamefont {S.}~\bibnamefont {Takahashi}},\
  }\href@noop {} {\bibfield  {journal} {\bibinfo  {journal} {Phys. Rev. B}\
  }\textbf {\bibinfo {volume} {50}},\ \bibinfo {pages} {7065} (\bibinfo {year}
  {1994})}\BibitemShut {NoStop}%
\bibitem [{\citenamefont {Koyama}\ and\ \citenamefont
  {Tachiki}(1995)}]{Koyama95}%
  \BibitemOpen
  \bibfield  {author} {\bibinfo {author} {\bibfnamefont {T.}~\bibnamefont
  {Koyama}}\ and\ \bibinfo {author} {\bibfnamefont {M.}~\bibnamefont
  {Tachiki}},\ }\href@noop {} {\bibfield  {journal} {\bibinfo  {journal} {Solid
  State Commun.}\ }\textbf {\bibinfo {volume} {96}},\ \bibinfo {pages} {367}
  (\bibinfo {year} {1995})}\BibitemShut {NoStop}%
\bibitem [{\citenamefont {Tachiki}\ \emph {et~al.}(2005)\citenamefont
  {Tachiki}, \citenamefont {Iizuka}, \citenamefont {Minami}, \citenamefont
  {Tejima},\ and\ \citenamefont {Nakamura}}]{Tachiki05}%
  \BibitemOpen
  \bibfield  {author} {\bibinfo {author} {\bibfnamefont {M.}~\bibnamefont
  {Tachiki}}, \bibinfo {author} {\bibfnamefont {M.}~\bibnamefont {Iizuka}},
  \bibinfo {author} {\bibfnamefont {K.}~\bibnamefont {Minami}}, \bibinfo
  {author} {\bibfnamefont {S.}~\bibnamefont {Tejima}}, \ and\ \bibinfo {author}
  {\bibfnamefont {H.}~\bibnamefont {Nakamura}},\ }\href@noop {} {\bibfield
  {journal} {\bibinfo  {journal} {Phys. Rev. B}\ }\textbf {\bibinfo {volume}
  {71}},\ \bibinfo {pages} {134515} (\bibinfo {year} {2005})}\BibitemShut
  {NoStop}%
\bibitem [{\citenamefont {Bulaevskii}\ and\ \citenamefont
  {Koshelev}(2006{\natexlab{b}})}]{Bulaevskii06}%
  \BibitemOpen
  \bibfield  {author} {\bibinfo {author} {\bibfnamefont {L.~N.}\ \bibnamefont
  {Bulaevskii}}\ and\ \bibinfo {author} {\bibfnamefont {A.~E.}\ \bibnamefont
  {Koshelev}},\ }\href@noop {} {\bibfield  {journal} {\bibinfo  {journal} {J.
  of Supercond. Novel Magn.}\ }\textbf {\bibinfo {volume} {19}},\ \bibinfo
  {pages} {349} (\bibinfo {year} {2006}{\natexlab{b}})}\BibitemShut {NoStop}%
\bibitem [{\citenamefont {Bulaevskii}\ and\ \citenamefont
  {Koshelev}(2007)}]{Bulaevskii07}%
  \BibitemOpen
  \bibfield  {author} {\bibinfo {author} {\bibfnamefont {L.~N.}\ \bibnamefont
  {Bulaevskii}}\ and\ \bibinfo {author} {\bibfnamefont {A.~E.}\ \bibnamefont
  {Koshelev}},\ }\href@noop {} {\bibfield  {journal} {\bibinfo  {journal}
  {Phys. Rev. Lett.}\ }\textbf {\bibinfo {volume} {99}},\ \bibinfo {pages}
  {057002} (\bibinfo {year} {2007})}\BibitemShut {NoStop}%
\bibitem [{\citenamefont {Lin}\ \emph {et~al.}(2008)\citenamefont {Lin},
  \citenamefont {Hu},\ and\ \citenamefont {Tachiki}}]{szlin08}%
  \BibitemOpen
  \bibfield  {author} {\bibinfo {author} {\bibfnamefont {S.~Z.}\ \bibnamefont
  {Lin}}, \bibinfo {author} {\bibfnamefont {X.}~\bibnamefont {Hu}}, \ and\
  \bibinfo {author} {\bibfnamefont {M.}~\bibnamefont {Tachiki}},\ }\href@noop
  {} {\bibfield  {journal} {\bibinfo  {journal} {Phys. Rev.}\ }\textbf
  {\bibinfo {volume} {B77}},\ \bibinfo {pages} {014507} (\bibinfo {year}
  {2008})}\BibitemShut {NoStop}%
\bibitem [{\citenamefont {Koshelev}\ and\ \citenamefont
  {Bulaevskii}(2008)}]{Koshelev08}%
  \BibitemOpen
  \bibfield  {author} {\bibinfo {author} {\bibfnamefont {A.~E.}\ \bibnamefont
  {Koshelev}}\ and\ \bibinfo {author} {\bibfnamefont {L.~N.}\ \bibnamefont
  {Bulaevskii}},\ }\href@noop {} {\bibfield  {journal} {\bibinfo  {journal}
  {Phys. Rev. B}\ }\textbf {\bibinfo {volume} {77}},\ \bibinfo {pages} {014530}
  (\bibinfo {year} {2008})}\BibitemShut {NoStop}%
\bibitem [{\citenamefont {Lin}\ and\ \citenamefont {Hu}(2008)}]{szlin08b}%
  \BibitemOpen
  \bibfield  {author} {\bibinfo {author} {\bibfnamefont {S.~Z.}\ \bibnamefont
  {Lin}}\ and\ \bibinfo {author} {\bibfnamefont {X.}~\bibnamefont {Hu}},\
  }\href@noop {} {\bibfield  {journal} {\bibinfo  {journal} {Phys. Rev. Lett.}\
  }\textbf {\bibinfo {volume} {100}},\ \bibinfo {pages} {247006} (\bibinfo
  {year} {2008})}\BibitemShut {NoStop}%
\bibitem [{\citenamefont {Tachiki}\ \emph {et~al.}(2009)\citenamefont
  {Tachiki}, \citenamefont {Fukuya},\ and\ \citenamefont {Koyama}}]{Tachiki09}%
  \BibitemOpen
  \bibfield  {author} {\bibinfo {author} {\bibfnamefont {M.}~\bibnamefont
  {Tachiki}}, \bibinfo {author} {\bibfnamefont {S.}~\bibnamefont {Fukuya}}, \
  and\ \bibinfo {author} {\bibfnamefont {T.}~\bibnamefont {Koyama}},\
  }\href@noop {} {\bibfield  {journal} {\bibinfo  {journal} {Phys. Rev. Lett.}\
  }\textbf {\bibinfo {volume} {102}},\ \bibinfo {pages} {127002} (\bibinfo
  {year} {2009})}\BibitemShut {NoStop}%
\bibitem [{\citenamefont {Lin}\ and\ \citenamefont {Hu}(2009)}]{szlin09a}%
  \BibitemOpen
  \bibfield  {author} {\bibinfo {author} {\bibfnamefont {S.~Z.}\ \bibnamefont
  {Lin}}\ and\ \bibinfo {author} {\bibfnamefont {X.}~\bibnamefont {Hu}},\
  }\href@noop {} {\bibfield  {journal} {\bibinfo  {journal} {Phys. Rev. B}\
  }\textbf {\bibinfo {volume} {79}},\ \bibinfo {pages} {104507} (\bibinfo
  {year} {2009})}\BibitemShut {NoStop}%
\bibitem [{\citenamefont {Wang}\ \emph {et~al.}(2009)\citenamefont {Wang},
  \citenamefont {Gu\'{e}non}, \citenamefont {Yuan}, \citenamefont {Iishi},
  \citenamefont {Arisawa}, \citenamefont {Hatano}, \citenamefont {Yamashita},
  \citenamefont {Koelle},\ and\ \citenamefont {Kleiner}}]{Wang09}%
  \BibitemOpen
  \bibfield  {author} {\bibinfo {author} {\bibfnamefont {H.~B.}\ \bibnamefont
  {Wang}}, \bibinfo {author} {\bibfnamefont {S.}~\bibnamefont {Gu\'{e}non}},
  \bibinfo {author} {\bibfnamefont {J.}~\bibnamefont {Yuan}}, \bibinfo {author}
  {\bibfnamefont {A.}~\bibnamefont {Iishi}}, \bibinfo {author} {\bibfnamefont
  {S.}~\bibnamefont {Arisawa}}, \bibinfo {author} {\bibfnamefont
  {T.}~\bibnamefont {Hatano}}, \bibinfo {author} {\bibfnamefont
  {T.}~\bibnamefont {Yamashita}}, \bibinfo {author} {\bibfnamefont
  {D.}~\bibnamefont {Koelle}}, \ and\ \bibinfo {author} {\bibfnamefont
  {R.}~\bibnamefont {Kleiner}},\ }\href@noop {} {\bibfield  {journal} {\bibinfo
   {journal} {Phys. Rev. Lett.}\ }\textbf {\bibinfo {volume} {102}},\ \bibinfo
  {pages} {017006} (\bibinfo {year} {2009})}\BibitemShut {NoStop}%
\bibitem [{\citenamefont {Rakhmanov}\ \emph {et~al.}(2009)\citenamefont
  {Rakhmanov}, \citenamefont {Savel'ev},\ and\ \citenamefont
  {Nori}}]{Rakhmanov09}%
  \BibitemOpen
  \bibfield  {author} {\bibinfo {author} {\bibfnamefont {A.~L.}\ \bibnamefont
  {Rakhmanov}}, \bibinfo {author} {\bibfnamefont {S.~E.}\ \bibnamefont
  {Savel'ev}}, \ and\ \bibinfo {author} {\bibfnamefont {F.}~\bibnamefont
  {Nori}},\ }\href@noop {} {\bibfield  {journal} {\bibinfo  {journal} {Phys.
  Rev. B}\ }\textbf {\bibinfo {volume} {79}},\ \bibinfo {pages} {184504}
  (\bibinfo {year} {2009})}\BibitemShut {NoStop}%
\bibitem [{\citenamefont {Wang}\ \emph {et~al.}(2010)\citenamefont {Wang},
  \citenamefont {Gu\'{e}non}, \citenamefont {Gross}, \citenamefont {Yuan},
  \citenamefont {Jiang}, \citenamefont {Zhong}, \citenamefont {Grunzweig},
  \citenamefont {Iishi}, \citenamefont {Wu}, \citenamefont {Hatano},
  \citenamefont {Koelle},\ and\ \citenamefont {Kleiner}}]{Wang10}%
  \BibitemOpen
  \bibfield  {author} {\bibinfo {author} {\bibfnamefont {H.~B.}\ \bibnamefont
  {Wang}}, \bibinfo {author} {\bibfnamefont {S.}~\bibnamefont {Gu\'{e}non}},
  \bibinfo {author} {\bibfnamefont {B.}~\bibnamefont {Gross}}, \bibinfo
  {author} {\bibfnamefont {J.}~\bibnamefont {Yuan}}, \bibinfo {author}
  {\bibfnamefont {Z.~G.}\ \bibnamefont {Jiang}}, \bibinfo {author}
  {\bibfnamefont {Y.~Y.}\ \bibnamefont {Zhong}}, \bibinfo {author}
  {\bibfnamefont {M.}~\bibnamefont {Grunzweig}}, \bibinfo {author}
  {\bibfnamefont {A.}~\bibnamefont {Iishi}}, \bibinfo {author} {\bibfnamefont
  {P.~H.}\ \bibnamefont {Wu}}, \bibinfo {author} {\bibfnamefont
  {T.}~\bibnamefont {Hatano}}, \bibinfo {author} {\bibfnamefont
  {D.}~\bibnamefont {Koelle}}, \ and\ \bibinfo {author} {\bibfnamefont
  {R.}~\bibnamefont {Kleiner}},\ }\href@noop {} {\bibfield  {journal} {\bibinfo
   {journal} {Phys. Rev. Lett.}\ }\textbf {\bibinfo {volume} {105}},\ \bibinfo
  {pages} {057002} (\bibinfo {year} {2010})}\BibitemShut {NoStop}%
\bibitem [{\citenamefont {Tsujimoto}\ \emph {et~al.}(2010)\citenamefont
  {Tsujimoto}, \citenamefont {Yamaki}, \citenamefont {Deguchi}, \citenamefont
  {Yamamoto}, \citenamefont {Kashiwagi}, \citenamefont {Minami}, \citenamefont
  {Tachiki}, \citenamefont {Kadowaki},\ and\ \citenamefont
  {Klemm}}]{Tsujimoto10}%
  \BibitemOpen
  \bibfield  {author} {\bibinfo {author} {\bibfnamefont {M.}~\bibnamefont
  {Tsujimoto}}, \bibinfo {author} {\bibfnamefont {K.}~\bibnamefont {Yamaki}},
  \bibinfo {author} {\bibfnamefont {K.}~\bibnamefont {Deguchi}}, \bibinfo
  {author} {\bibfnamefont {T.}~\bibnamefont {Yamamoto}}, \bibinfo {author}
  {\bibfnamefont {T.}~\bibnamefont {Kashiwagi}}, \bibinfo {author}
  {\bibfnamefont {H.}~\bibnamefont {Minami}}, \bibinfo {author} {\bibfnamefont
  {M.}~\bibnamefont {Tachiki}}, \bibinfo {author} {\bibfnamefont
  {K.}~\bibnamefont {Kadowaki}}, \ and\ \bibinfo {author} {\bibfnamefont
  {R.~A.}\ \bibnamefont {Klemm}},\ }\href@noop {} {\bibfield  {journal}
  {\bibinfo  {journal} {Phys. Rev. Lett.}\ }\textbf {\bibinfo {volume} {105}},\
  \bibinfo {pages} {037005} (\bibinfo {year} {2010})}\BibitemShut {NoStop}%
\bibitem [{\citenamefont {Tsujimoto}\ \emph {et~al.}(2012)\citenamefont
  {Tsujimoto}, \citenamefont {Yamamoto}, \citenamefont {Delfanazari},
  \citenamefont {Nakayama}, \citenamefont {Kitamura}, \citenamefont {Sawamura},
  \citenamefont {Kashiwagi}, \citenamefont {Minami}, \citenamefont {Tachiki},
  \citenamefont {Kadowaki},\ and\ \citenamefont {Klemm}}]{Tsujimoto12b}%
  \BibitemOpen
  \bibfield  {author} {\bibinfo {author} {\bibfnamefont {M.}~\bibnamefont
  {Tsujimoto}}, \bibinfo {author} {\bibfnamefont {T.}~\bibnamefont {Yamamoto}},
  \bibinfo {author} {\bibfnamefont {K.}~\bibnamefont {Delfanazari}}, \bibinfo
  {author} {\bibfnamefont {R.}~\bibnamefont {Nakayama}}, \bibinfo {author}
  {\bibfnamefont {T.}~\bibnamefont {Kitamura}}, \bibinfo {author}
  {\bibfnamefont {M.}~\bibnamefont {Sawamura}}, \bibinfo {author}
  {\bibfnamefont {T.}~\bibnamefont {Kashiwagi}}, \bibinfo {author}
  {\bibfnamefont {H.}~\bibnamefont {Minami}}, \bibinfo {author} {\bibfnamefont
  {M.}~\bibnamefont {Tachiki}}, \bibinfo {author} {\bibfnamefont
  {K.}~\bibnamefont {Kadowaki}}, \ and\ \bibinfo {author} {\bibfnamefont
  {R.~A.}\ \bibnamefont {Klemm}},\ }\href@noop {} {\bibfield  {journal}
  {\bibinfo  {journal} {Phys. Rev. Lett.}\ }\textbf {\bibinfo {volume} {108}},\
  \bibinfo {pages} {107006} (\bibinfo {year} {2012})}\BibitemShut {NoStop}%
\bibitem [{\citenamefont {Ozyuzer}\ \emph {et~al.}(2007)\citenamefont
  {Ozyuzer}, \citenamefont {Koshelev}, \citenamefont {Kurter}, \citenamefont
  {Gopalsami}, \citenamefont {Li}, \citenamefont {Tachiki}, \citenamefont
  {Kadowaki}, \citenamefont {Yamamoto}, \citenamefont {Minami}, \citenamefont
  {Yamaguchi}, \citenamefont {Tachiki}, \citenamefont {Gray}, \citenamefont
  {Kwok},\ and\ \citenamefont {Welp}}]{Ozyuzer07}%
  \BibitemOpen
  \bibfield  {author} {\bibinfo {author} {\bibfnamefont {L.}~\bibnamefont
  {Ozyuzer}}, \bibinfo {author} {\bibfnamefont {A.~E.}\ \bibnamefont
  {Koshelev}}, \bibinfo {author} {\bibfnamefont {C.}~\bibnamefont {Kurter}},
  \bibinfo {author} {\bibfnamefont {N.}~\bibnamefont {Gopalsami}}, \bibinfo
  {author} {\bibfnamefont {Q.}~\bibnamefont {Li}}, \bibinfo {author}
  {\bibfnamefont {M.}~\bibnamefont {Tachiki}}, \bibinfo {author} {\bibfnamefont
  {K.}~\bibnamefont {Kadowaki}}, \bibinfo {author} {\bibfnamefont
  {T.}~\bibnamefont {Yamamoto}}, \bibinfo {author} {\bibfnamefont
  {H.}~\bibnamefont {Minami}}, \bibinfo {author} {\bibfnamefont
  {H.}~\bibnamefont {Yamaguchi}}, \bibinfo {author} {\bibfnamefont
  {T.}~\bibnamefont {Tachiki}}, \bibinfo {author} {\bibfnamefont {K.~E.}\
  \bibnamefont {Gray}}, \bibinfo {author} {\bibfnamefont {W.~K.}\ \bibnamefont
  {Kwok}}, \ and\ \bibinfo {author} {\bibfnamefont {U.}~\bibnamefont {Welp}},\
  }\href@noop {} {\bibfield  {journal} {\bibinfo  {journal} {Science}\ }\textbf
  {\bibinfo {volume} {318}},\ \bibinfo {pages} {1291} (\bibinfo {year}
  {2007})}\BibitemShut {NoStop}%
\bibitem [{\citenamefont {Kashiwagi}\ \emph {et~al.}(2012)\citenamefont
  {Kashiwagi}, \citenamefont {Tsujimoto}, \citenamefont {Yamamoto},
  \citenamefont {Minami}, \citenamefont {Yamaki}, \citenamefont {Delfanazari},
  \citenamefont {Deguchi}, \citenamefont {Orita}, \citenamefont {Koike},
  \citenamefont {Nakayama}, \citenamefont {Kitamura}, \citenamefont {Sawamura},
  \citenamefont {Hagino}, \citenamefont {Ishida}, \citenamefont {Ivanovic},
  \citenamefont {Asai}, \citenamefont {Tachiki}, \citenamefont {Klemm},\ and\
  \citenamefont {Kadowaki}}]{Kashiwagi2012}%
  \BibitemOpen
  \bibfield  {author} {\bibinfo {author} {\bibfnamefont {T.}~\bibnamefont
  {Kashiwagi}}, \bibinfo {author} {\bibfnamefont {M.}~\bibnamefont
  {Tsujimoto}}, \bibinfo {author} {\bibfnamefont {T.}~\bibnamefont {Yamamoto}},
  \bibinfo {author} {\bibfnamefont {H.}~\bibnamefont {Minami}}, \bibinfo
  {author} {\bibfnamefont {K.}~\bibnamefont {Yamaki}}, \bibinfo {author}
  {\bibfnamefont {K.}~\bibnamefont {Delfanazari}}, \bibinfo {author}
  {\bibfnamefont {K.}~\bibnamefont {Deguchi}}, \bibinfo {author} {\bibfnamefont
  {N.}~\bibnamefont {Orita}}, \bibinfo {author} {\bibfnamefont
  {T.}~\bibnamefont {Koike}}, \bibinfo {author} {\bibfnamefont
  {R.}~\bibnamefont {Nakayama}}, \bibinfo {author} {\bibfnamefont
  {T.}~\bibnamefont {Kitamura}}, \bibinfo {author} {\bibfnamefont
  {M.}~\bibnamefont {Sawamura}}, \bibinfo {author} {\bibfnamefont
  {S.}~\bibnamefont {Hagino}}, \bibinfo {author} {\bibfnamefont
  {K.}~\bibnamefont {Ishida}}, \bibinfo {author} {\bibfnamefont
  {K.}~\bibnamefont {Ivanovic}}, \bibinfo {author} {\bibfnamefont
  {H.}~\bibnamefont {Asai}}, \bibinfo {author} {\bibfnamefont {M.}~\bibnamefont
  {Tachiki}}, \bibinfo {author} {\bibfnamefont {R.~A.}\ \bibnamefont {Klemm}},
  \ and\ \bibinfo {author} {\bibfnamefont {K.}~\bibnamefont {Kadowaki}},\
  }\href {\doibase 10.1143/JJAP.51.010113} {\bibfield  {journal} {\bibinfo
  {journal} {Jpn. J. Appl. Phys.}\ }\textbf {\bibinfo {volume} {51}},\ \bibinfo
  {pages} {010113} (\bibinfo {year} {2012})}\BibitemShut {NoStop}%
\bibitem [{\citenamefont {Koshelev}(2010)}]{Koshelev10}%
  \BibitemOpen
  \bibfield  {author} {\bibinfo {author} {\bibfnamefont {A.~E.}\ \bibnamefont
  {Koshelev}},\ }\href {\doibase 10.1103/PhysRevB.82.174512} {\bibfield
  {journal} {\bibinfo  {journal} {Phys. Rev. B}\ }\textbf {\bibinfo {volume}
  {82}},\ \bibinfo {pages} {174512} (\bibinfo {year} {2010})}\BibitemShut
  {NoStop}%
\bibitem [{\citenamefont {Lin}\ and\ \citenamefont {Hu}(2012)}]{szlin12a}%
  \BibitemOpen
  \bibfield  {author} {\bibinfo {author} {\bibfnamefont {S.-Z.}\ \bibnamefont
  {Lin}}\ and\ \bibinfo {author} {\bibfnamefont {X.}~\bibnamefont {Hu}},\
  }\href {\doibase 10.1103/PhysRevB.86.054506} {\bibfield  {journal} {\bibinfo
  {journal} {Phys. Rev. B}\ }\textbf {\bibinfo {volume} {86}},\ \bibinfo
  {pages} {054506} (\bibinfo {year} {2012})}\BibitemShut {NoStop}%
\bibitem [{\citenamefont {Hu}\ and\ \citenamefont {Lin}(2010)}]{Hu10}%
  \BibitemOpen
  \bibfield  {author} {\bibinfo {author} {\bibfnamefont {X.}~\bibnamefont
  {Hu}}\ and\ \bibinfo {author} {\bibfnamefont {S.~Z.}\ \bibnamefont {Lin}},\
  }\href@noop {} {\bibfield  {journal} {\bibinfo  {journal} {Supercond. Sci.
  Technol.}\ }\textbf {\bibinfo {volume} {23}},\ \bibinfo {pages} {053001}
  (\bibinfo {year} {2010})}\BibitemShut {NoStop}%
\bibitem [{\citenamefont {Savel'ev}\ \emph {et~al.}(2010)\citenamefont
  {Savel'ev}, \citenamefont {Yampol'skii}, \citenamefont {Rakhmanov},\ and\
  \citenamefont {Nori}}]{Savelev10}%
  \BibitemOpen
  \bibfield  {author} {\bibinfo {author} {\bibfnamefont {S.}~\bibnamefont
  {Savel'ev}}, \bibinfo {author} {\bibfnamefont {V.~A.}\ \bibnamefont
  {Yampol'skii}}, \bibinfo {author} {\bibfnamefont {A.~L.}\ \bibnamefont
  {Rakhmanov}}, \ and\ \bibinfo {author} {\bibfnamefont {F.}~\bibnamefont
  {Nori}},\ }\href@noop {} {\bibfield  {journal} {\bibinfo  {journal} {Rep.
  Prog. Phys.}\ }\textbf {\bibinfo {volume} {73}},\ \bibinfo {pages} {026501}
  (\bibinfo {year} {2010})}\BibitemShut {NoStop}%
\bibitem [{\citenamefont {Welp}\ \emph {et~al.}(2013)\citenamefont {Welp},
  \citenamefont {Kadowaki},\ and\ \citenamefont {Kleiner}}]{Welp13}%
  \BibitemOpen
  \bibfield  {author} {\bibinfo {author} {\bibfnamefont {U.}~\bibnamefont
  {Welp}}, \bibinfo {author} {\bibfnamefont {K.}~\bibnamefont {Kadowaki}}, \
  and\ \bibinfo {author} {\bibfnamefont {R.}~\bibnamefont {Kleiner}},\ }\href
  {\doibase 10.1038/nphoton.2013.216} {\bibfield  {journal} {\bibinfo
  {journal} {Nat Photon}\ }\textbf {\bibinfo {volume} {7}},\ \bibinfo {pages}
  {702} (\bibinfo {year} {2013})}\BibitemShut {NoStop}%
\bibitem [{\citenamefont {Li}\ \emph {et~al.}(2012)\citenamefont {Li},
  \citenamefont {Yuan}, \citenamefont {Kinev}, \citenamefont {Li},
  \citenamefont {Gross}, \citenamefont {Gu\'enon}, \citenamefont {Ishii},
  \citenamefont {Hirata}, \citenamefont {Hatano}, \citenamefont {Koelle},
  \citenamefont {Kleiner}, \citenamefont {Koshelets}, \citenamefont {Wang},\
  and\ \citenamefont {Wu}}]{LiMengyue12}%
  \BibitemOpen
  \bibfield  {author} {\bibinfo {author} {\bibfnamefont {M.}~\bibnamefont
  {Li}}, \bibinfo {author} {\bibfnamefont {J.}~\bibnamefont {Yuan}}, \bibinfo
  {author} {\bibfnamefont {N.}~\bibnamefont {Kinev}}, \bibinfo {author}
  {\bibfnamefont {J.}~\bibnamefont {Li}}, \bibinfo {author} {\bibfnamefont
  {B.}~\bibnamefont {Gross}}, \bibinfo {author} {\bibfnamefont
  {S.}~\bibnamefont {Gu\'enon}}, \bibinfo {author} {\bibfnamefont
  {A.}~\bibnamefont {Ishii}}, \bibinfo {author} {\bibfnamefont
  {K.}~\bibnamefont {Hirata}}, \bibinfo {author} {\bibfnamefont
  {T.}~\bibnamefont {Hatano}}, \bibinfo {author} {\bibfnamefont
  {D.}~\bibnamefont {Koelle}}, \bibinfo {author} {\bibfnamefont
  {R.}~\bibnamefont {Kleiner}}, \bibinfo {author} {\bibfnamefont {V.~P.}\
  \bibnamefont {Koshelets}}, \bibinfo {author} {\bibfnamefont {H.}~\bibnamefont
  {Wang}}, \ and\ \bibinfo {author} {\bibfnamefont {P.}~\bibnamefont {Wu}},\
  }\href {\doibase 10.1103/PhysRevB.86.060505} {\bibfield  {journal} {\bibinfo
  {journal} {Phys. Rev. B}\ }\textbf {\bibinfo {volume} {86}},\ \bibinfo
  {pages} {060505} (\bibinfo {year} {2012})}\BibitemShut {NoStop}%
\bibitem [{\citenamefont {Lin}\ and\ \citenamefont
  {Koshelev}(2013{\natexlab{a}})}]{szlin13b}%
  \BibitemOpen
  \bibfield  {author} {\bibinfo {author} {\bibfnamefont {S.-Z.}\ \bibnamefont
  {Lin}}\ and\ \bibinfo {author} {\bibfnamefont {A.~E.}\ \bibnamefont
  {Koshelev}},\ }\href {\doibase 10.1103/PhysRevB.87.214511} {\bibfield
  {journal} {\bibinfo  {journal} {Phys. Rev. B}\ }\textbf {\bibinfo {volume}
  {87}},\ \bibinfo {pages} {214511} (\bibinfo {year}
  {2013}{\natexlab{a}})}\BibitemShut {NoStop}%
\bibitem [{\citenamefont {Bulaevskii}\ \emph {et~al.}(2011)\citenamefont
  {Bulaevskii}, \citenamefont {Martin},\ and\ \citenamefont
  {Hal\'asz}}]{Bulaevskii11}%
  \BibitemOpen
  \bibfield  {author} {\bibinfo {author} {\bibfnamefont {L.~N.}\ \bibnamefont
  {Bulaevskii}}, \bibinfo {author} {\bibfnamefont {I.}~\bibnamefont {Martin}},
  \ and\ \bibinfo {author} {\bibfnamefont {G.~B.}\ \bibnamefont {Hal\'asz}},\
  }\href {\doibase 10.1103/PhysRevB.84.014516} {\bibfield  {journal} {\bibinfo
  {journal} {Phys. Rev. B}\ }\textbf {\bibinfo {volume} {84}},\ \bibinfo
  {pages} {014516} (\bibinfo {year} {2011})}\BibitemShut {NoStop}%
\bibitem [{\citenamefont {Lin}\ \emph {et~al.}(2011)\citenamefont {Lin},
  \citenamefont {Hu},\ and\ \citenamefont {Bulaevskii}}]{szlin11}%
  \BibitemOpen
  \bibfield  {author} {\bibinfo {author} {\bibfnamefont {S.-Z.}\ \bibnamefont
  {Lin}}, \bibinfo {author} {\bibfnamefont {X.}~\bibnamefont {Hu}}, \ and\
  \bibinfo {author} {\bibfnamefont {L.}~\bibnamefont {Bulaevskii}},\ }\href
  {\doibase 10.1103/PhysRevB.84.104501} {\bibfield  {journal} {\bibinfo
  {journal} {Phys. Rev. B}\ }\textbf {\bibinfo {volume} {84}},\ \bibinfo
  {pages} {104501} (\bibinfo {year} {2011})}\BibitemShut {NoStop}%
\bibitem [{\citenamefont {Hadley}\ \emph {et~al.}(1988)\citenamefont {Hadley},
  \citenamefont {Beasley},\ and\ \citenamefont {Wiesenfeld}}]{Hadley88}%
  \BibitemOpen
  \bibfield  {author} {\bibinfo {author} {\bibfnamefont {P.}~\bibnamefont
  {Hadley}}, \bibinfo {author} {\bibfnamefont {M.~R.}\ \bibnamefont {Beasley}},
  \ and\ \bibinfo {author} {\bibfnamefont {K.}~\bibnamefont {Wiesenfeld}},\
  }\href {\doibase 10.1103/PhysRevB.38.8712} {\bibfield  {journal} {\bibinfo
  {journal} {Phys. Rev. B}\ }\textbf {\bibinfo {volume} {38}},\ \bibinfo
  {pages} {8712} (\bibinfo {year} {1988})}\BibitemShut {NoStop}%
\bibitem [{\citenamefont {Koyama}\ and\ \citenamefont
  {Tachiki}(1996)}]{Koyama96}%
  \BibitemOpen
  \bibfield  {author} {\bibinfo {author} {\bibfnamefont {T.}~\bibnamefont
  {Koyama}}\ and\ \bibinfo {author} {\bibfnamefont {M.}~\bibnamefont
  {Tachiki}},\ }\href@noop {} {\bibfield  {journal} {\bibinfo  {journal} {Phys.
  Rev. B}\ }\textbf {\bibinfo {volume} {54}},\ \bibinfo {pages} {16183}
  (\bibinfo {year} {1996})}\BibitemShut {NoStop}%
\bibitem [{\citenamefont {Acebr\'on}\ \emph {et~al.}(2005)\citenamefont
  {Acebr\'on}, \citenamefont {Bonilla}, \citenamefont {P\'erez~Vicente},
  \citenamefont {Ritort},\ and\ \citenamefont {Spigler}}]{Acebron05}%
  \BibitemOpen
  \bibfield  {author} {\bibinfo {author} {\bibfnamefont {J.~A.}\ \bibnamefont
  {Acebr\'on}}, \bibinfo {author} {\bibfnamefont {L.~L.}\ \bibnamefont
  {Bonilla}}, \bibinfo {author} {\bibfnamefont {C.~J.}\ \bibnamefont
  {P\'erez~Vicente}}, \bibinfo {author} {\bibfnamefont {F.}~\bibnamefont
  {Ritort}}, \ and\ \bibinfo {author} {\bibfnamefont {R.}~\bibnamefont
  {Spigler}},\ }\href {\doibase 10.1103/RevModPhys.77.137} {\bibfield
  {journal} {\bibinfo  {journal} {Rev. Mod. Phys.}\ }\textbf {\bibinfo {volume}
  {77}},\ \bibinfo {pages} {137} (\bibinfo {year} {2005})}\BibitemShut
  {NoStop}%
\bibitem [{\citenamefont {Orita}\ \emph {et~al.}(2010)\citenamefont {Orita},
  \citenamefont {Minami}, \citenamefont {Koike}, \citenamefont {Yamamoto},\
  and\ \citenamefont {Kadowaki}}]{Orita2010}%
  \BibitemOpen
  \bibfield  {author} {\bibinfo {author} {\bibfnamefont {N.}~\bibnamefont
  {Orita}}, \bibinfo {author} {\bibfnamefont {H.}~\bibnamefont {Minami}},
  \bibinfo {author} {\bibfnamefont {T.}~\bibnamefont {Koike}}, \bibinfo
  {author} {\bibfnamefont {T.}~\bibnamefont {Yamamoto}}, \ and\ \bibinfo
  {author} {\bibfnamefont {K.}~\bibnamefont {Kadowaki}},\ }\href@noop {}
  {\bibfield  {journal} {\bibinfo  {journal} {Physica C}\ }\textbf {\bibinfo
  {volume} {470}},\ \bibinfo {pages} {S786} (\bibinfo {year}
  {2010})}\BibitemShut {NoStop}%
\bibitem [{\citenamefont {Benseman}\ \emph {et~al.}(2013)\citenamefont
  {Benseman}, \citenamefont {Gray}, \citenamefont {Koshelev}, \citenamefont
  {Kwok}, \citenamefont {Welp}, \citenamefont {Minami}, \citenamefont
  {Kadowaki},\ and\ \citenamefont {Yamamoto}}]{Benseman12}%
  \BibitemOpen
  \bibfield  {author} {\bibinfo {author} {\bibfnamefont {T.~M.}\ \bibnamefont
  {Benseman}}, \bibinfo {author} {\bibfnamefont {K.~E.}\ \bibnamefont {Gray}},
  \bibinfo {author} {\bibfnamefont {A.~E.}\ \bibnamefont {Koshelev}}, \bibinfo
  {author} {\bibfnamefont {W.-K.}\ \bibnamefont {Kwok}}, \bibinfo {author}
  {\bibfnamefont {U.}~\bibnamefont {Welp}}, \bibinfo {author} {\bibfnamefont
  {H.}~\bibnamefont {Minami}}, \bibinfo {author} {\bibfnamefont
  {K.}~\bibnamefont {Kadowaki}}, \ and\ \bibinfo {author} {\bibfnamefont
  {T.}~\bibnamefont {Yamamoto}},\ }\href {\doibase
  http://dx.doi.org/10.1063/1.4813536} {\bibfield  {journal} {\bibinfo
  {journal} {Appl. Phys. Lett.}\ }\textbf {\bibinfo {volume} {103}},\ \bibinfo
  {eid} {022602} (\bibinfo {year} {2013})}\BibitemShut {NoStop}%
\bibitem [{\citenamefont {Lin}\ and\ \citenamefont
  {Koshelev}(2013{\natexlab{b}})}]{szlin13}%
  \BibitemOpen
  \bibfield  {author} {\bibinfo {author} {\bibfnamefont {S.-Z.}\ \bibnamefont
  {Lin}}\ and\ \bibinfo {author} {\bibfnamefont {A.~E.}\ \bibnamefont
  {Koshelev}},\ }\href {\doibase 10.1016/j.physc.2012.11.008} {\bibfield
  {journal} {\bibinfo  {journal} {Physica C}\ }\textbf {\bibinfo {volume}
  {491}},\ \bibinfo {pages} {24} (\bibinfo {year}
  {2013}{\natexlab{b}})}\BibitemShut {NoStop}%
\bibitem [{\citenamefont {Hong}\ \emph {et~al.}(2005)\citenamefont {Hong},
  \citenamefont {Park},\ and\ \citenamefont {Choi}}]{Hong05}%
  \BibitemOpen
  \bibfield  {author} {\bibinfo {author} {\bibfnamefont {H.}~\bibnamefont
  {Hong}}, \bibinfo {author} {\bibfnamefont {H.}~\bibnamefont {Park}}, \ and\
  \bibinfo {author} {\bibfnamefont {M.~Y.}\ \bibnamefont {Choi}},\ }\href
  {\doibase 10.1103/PhysRevE.72.036217} {\bibfield  {journal} {\bibinfo
  {journal} {Phys. Rev. E}\ }\textbf {\bibinfo {volume} {72}},\ \bibinfo
  {pages} {036217} (\bibinfo {year} {2005})}\BibitemShut {NoStop}%
\end{thebibliography}
%

\end{document}